\documentclass[12pt]{article}

\usepackage{cite}
\usepackage{epsfig}
\usepackage{graphicx}
\usepackage{amsmath}
\usepackage{amssymb}
\usepackage{ulem}
\usepackage{mdwlist}          
\usepackage{color}              

\usepackage{a41}
\usepackage{color}
\usepackage[rflt]{floatflt}
\usepackage{float}
\usepackage{slashed}

\usepackage{calrsfs}
\DeclareMathAlphabet{\pazocal}{OMS}{zplm}{m}{n}
\newcommand{\Ca}{\mathcal{C}}

\newcommand{\Da}{\mathcal{D}}

\usepackage{array}

\usepackage[english]{babel}
\usepackage[latin1]{inputenc}
\usepackage[T1]{fontenc}
\usepackage{ae}

\usepackage{url}

\usepackage{amsmath, amsthm, amssymb}
\newtheorem{thm}{Theorem}[section]

\newtheorem{definition}[thm]{Definition}

 \newcommand{\GeV}{\mathrm{GeV}}

\newcommand{\Mvec}{{\rm\bf M}}

\usepackage{rotating}

\usepackage{graphicx}

\newcounter{mmacnt}
\def\restartmma{\setcounter{mmacnt}{0}}
\restartmma \catcode`|=\active
\def|#1|{\mathrm{#1}}
\catcode`|=12
\newenvironment{mma}{
 \par\smallskip
 \catcode`|=\active
 \parskip=0pt\parindent=0pt 
 \small
 \def\In##1\\{%
   \def\linebreak{\hfill\break\null\qquad}%
   \refstepcounter{mmacnt}
   \hangindent=2.5em\hangafter=0
   \leavevmode
   \llap{\tiny\sffamily In[\arabic{mmacnt}]:=\kern.5em}%
   \mathversion{bold}\footnotesize$\displaystyle##1$\normalsize
   \mathversion{normal}\par
 }%
 \def\\Print##1\\{%
   \def\linebreak{		\hfill\break}%
   \hangindent=2.5em\hangafter=0
   \leavevmode ##1\par}%
 \def\Out##1\\{%
   \def\linebreak{$\hfill\break\null\hfill$}%
   \kern\abovedisplayskip\par
   \hangindent=2.5em\hangafter=0
   \leavevmode
   \llap{\tiny\sffamily Out[\arabic{mmacnt}]=\kern.5em}
   \footnotesize$\displaystyle##1$\normalsize\hfill\null\par
   \kern\belowdisplayskip
 }%
 \def\Warning##1##2\\{%
   \def\linebreak{\hfill\break}%
   \hangindent=2.5em\hangafter=0
   \leavevmode
   {\scriptsize##1 : ##2}\par}%
}{%
 \par\smallskip
}

\usepackage{color}

\newenvironment{fshaded}{%
\MakeFramed {\FrameRestore}
}%
{\endMakeFramed}

\allowdisplaybreaks[4]

\begin{document}
\setlength{\baselineskip}{0.515cm}
\sloppy
\thispagestyle{empty}
\begin{flushleft}
DESY 24--037
\hfill 
\\
ZU-TH 58/25
\\
RISC Report series 25-04 \hfill September 2025
\\
MPP-2025-189
\\
\end{flushleft}

\mbox{}
\vspace*{\fill}
\begin{center}

{\LARGE\bf The Single-Mass Variable Flavor Number Scheme at} 

\vspace*{3mm} 
{\LARGE\bf Three-Loop Order}

\vspace{3cm}
\large
J.~Ablinger$^{a}$,
A.~Behring$^b$, 
J.~Bl\"umlein$^{c,d}$, 
A.~De Freitas$^{a}$, \\
A.~von Manteuffel$^e$, 
C.~Schneider$^a$, 
and
K.~Sch\"onwald$^f$\footnote{Present address: CERN, Theoretical Physics Department, CH-1211 
Geneva 23, Switzerland.}

\vspace{1.cm}
\normalsize
{\it $^a$~Research Institute for Symbolic Computation (RISC),\\
                          Johannes Kepler University Linz, Altenbergerstra\ss{}e 69,
                          A--4040 Linz, Austria}

\vspace*{2mm}
{\it $^b$~Max-Planck-Institut f\"ur Physik,
Boltzmannstra\ss{}e 8, 85748 Garching, Germany}

\vspace*{2mm}
{\it $^c$ Deutsches Elektronen-Synchrotron DESY, Platanenallee 6, 15738 Zeuthen, Germany} 

\vspace*{2mm}
{\it $^d$ Institut f\"ur Theoretische Physik III, IV, TU Dortmund, Otto-Hahn Stra\ss{}e 4, \newline 
44227 Dortmund, 
Germany}

\vspace*{2mm}
{\it $^e$~Institut f\"ur Theoretische Physik, Universit\"at Regensburg, 93040 Regensburg, 
Germany} 

\vspace*{2mm}
{\it $^f$~Physik-Institut, Universit\"at Z\"urich, Winterthurerstra\ss{}e 190, CH-8057 Z\"urich, 
Switzerland}

\vspace*{3mm}


\end{center}
\normalsize
\vspace{\fill}
\begin{abstract}
\noindent
The matching relations in the unpolarized and polarized variable flavor number scheme at 
three-loop order are presented in the single-mass case. They describe the process of massive 
quarks becoming light at large virtualities $Q^2$. In this framework, heavy-quark parton 
distributions can be defined. Numerical results are presented on the matching relations in 
the case of the single-mass variable flavor number scheme for the light parton, charm 
and bottom quark distributions. These relations are process independent. In the polarized 
case we generally work in the Larin scheme. To two-loop order we present the polarized 
massive OMEs also in the $\overline{\rm MS}$ scheme. Fast numerical codes for the 
single-mass massive operator matrix elements are provided. 
\end{abstract}

\vspace*{\fill}
\noindent

\newpage

\vspace*{1mm}
\noindent
\section{Introduction}
\label{sec:1}

\vspace*{1mm}
\noindent
The massive operator matrix elements (OMEs) \cite{Buza:1995ie,Ablinger:2017err} in Quantum 
Chromodynamics (QCD) are process-independent quantities. By virtue of the renormalization 
group equations they allow to describe the transition of heavy-quark contributions to become 
light at high virtualities $Q^2$ in deep-inelastic scattering 
\cite{Politzer:1974fr,Buras:1979yt,Reya:1979zk,Blumlein:2012bf,Blumlein:2023aso} and other hard 
scattering 
processes \cite{Chen:2022tpk}. In this way, also so-called heavy-flavor parton distributions 
can be defined in the single-mass \cite{Buza:1996wv,Bierenbaum:2009mv} and 
two-mass \cite{Ablinger:2017err} cases. Originally, perturbative calculations of the 
heavy-flavor corrections to structure functions were performed in the so-called fixed flavor 
number scheme (FFNS), see Refs.~\cite{Gluck:1980cp,Gluck:1987uk}, adding the heavy-flavor Wilson 
coefficients to the massless ones. Here one has also to account for virtual corrections, 
i.e., a part of the heavy-flavor contributions may have purely massless final states but 
virtual heavy-flavor lines.

The idea of the variable flavor number scheme (VFNS) is now to establish the transition of a 
generally massive quark to act as a massless one at high scales of the virtuality $Q^2$ in 
the deep-inelastic scattering process. This is achieved by the relation \cite{Buza:1996wv}
\begin{eqnarray}
F_i(N_F,Q^2) + \lim_{Q^2 \gg m^2_Q} F_{i,H}(N_F,Q^2,m^2_Q) = F_i(N_F+1,Q^2),
\label{eq1}
\end{eqnarray}
for the transition of $N_F$ to $N_F+1$ light flavors. Here $m_Q$ denotes the heavy quark 
mass, $F_i$ 
is the light-flavor 
part of the structure function and $F_{i,H}$ denotes all its heavy-flavor 
parts. Equation~(\ref{eq1}) is valid in regions where power corrections 
$(m^2_Q/Q^2)^k, k \geq 1$ can be safely neglected. This relation has to obey the renormalization 
group 
equations \cite{Callan:1970yg,Symanzik:1970rt}, which imply the corresponding matching relations 
of the different parton distribution functions (PDFs) via the massive operator matrix 
elements 
(OMEs).\footnote{In the literature also modifications of the single-mass VFNS, Eq.~(\ref{eq1}), 
such 
as the ACOT scheme \cite{Aivazis:1993pi,Collins:1998rz} and the FONLL scheme 
\cite{Cacciari:1998it,Forte:2010ta} have been discussed, to one- and two-loop order. Aspects of 
the ACOT scheme, in comparison to the present approach, have been discussed in 
Ref.~\cite{Buza:1996wv} in detail already. The FONLL scheme builds on the massive OMEs 
calculated 
in the approach of Refs.~\cite{Buza:1995ie,Bierenbaum:2009mv}.} At next-to-leading-order 
(NLO) it was 
shown  \cite{Gluck:1993dpa} that these logarithms emerging in the FFNS in the case of the charm 
quark corrections show a rather stable behavior under the variation of the factorization and 
renormalization scales within the kinematic region at HERA. This may become different at even 
higher virtualities, requiring their resummation at these scales in the VFNS. 

The building blocks to set up the VFNS are the massive OMEs. In the unpolarized and polarized 
cases they have been computed to two-loop order\footnote{Some of the one-loop OMEs can be 
extracted from the results of 
Refs.~\cite{Witten:1975bh,Babcock:1977fi,Shifman:1977yb,Leveille:1978px,Watson:1981ce}.}
in Refs.~\cite{Buza:1995ie,Buza:1996wv,Buza:1996xr,Bierenbaum:2007qe,Bierenbaum:2009zt,
Behring:2014eya,Watson:1981ce,Bierenbaum:2007pn,Blumlein:2016xcy,Hekhorn:2018ywm,
Blumlein:2019qze,Blumlein:2019zux,Blumlein:2021xlc,Bierenbaum:2022biv}. 
A series of Mellin moments at three-loop order has been calculated in 
Ref.~\cite{Bierenbaum:2009mv}. From three-loop order on, seven unpolarized massive OMEs, 
$A_{qq,Q}^{\rm NS,+}, 
A_{Qq}^{\rm PS}, A_{qq,Q}^{\rm PS}, A_{Qg}, A_{qg,Q}, A_{gq,Q}$ and $A_{gg,Q}$, and the 
corresponding polarized 
OMEs contribute. Here $A_{qq,Q}^{\rm NS,+}$ appears in the transition relations for the combination
of the parton distributions $f_i + f_{\bar{i}}$, while $A_{qq,Q}^{\rm NS,-}$ rules the transition for
$f_i - f_{\bar{i}}$. The latter OME does not contribute to the structure functions $F_2(x,Q^2)$ or 
$g_1(x,Q^2)$, but stems from heavy-flavor corrections to the structure functions $xF_3(x,Q^2)$ 
or $g_4(x,Q^2)$, 
see Ref.~\cite{Blumlein:1996vs}.
For general values of the Mellin variable $N$ and the momentum fraction variable $z$ they were 
calculated in 
Refs.~\cite{Behring:2014eya,Ablinger:2010ty,Ablinger:2014lka,Ablinger:2014vwa,Ablinger:2014nga,
Ablinger:2022wbb,Ablinger:2023ahe,Ablinger:2024xtt,Blumlein:2016xcy,
Behring:2015zaa,Ablinger:2019etw,Behring:2021asx,
Blumlein:2021xlc}. 

Starting at two-loop order, there are also two-mass contributions. While they
have a factorizable structure at two-loop order \cite{Blumlein:2018jfm,Bierenbaum:2022biv}, they 
result also from irreducible 
Feynman diagrams containing two lines of different heavy-quark mass from three-loop order. In 
Ref.~\cite{Bierenbaum:2009zt} we corrected some of the two-loop massive OMEs of 
Ref.~\cite{Buza:1996wv}. Also 
the two-loop non-singlet (NS) results of Ref.~\cite{Buza:1995ie} had to be corrected, 
cf.~Ref.~\cite{Ablinger:2014lka}, 
since (inclusive) structure functions are considered. The five two-mass OMEs $\tilde{A}_{qq,Q}^{\rm NS}, 
\tilde{A}_{Qq}^{\rm PS}, \tilde{A}_{Qg}, \tilde{A}_{gq,Q}$ and $\tilde{A}_{gg,Q}$ have been calculated to 
three-loop order in Refs.~\cite{Ablinger:2017err,Ablinger:2017xml,
Ablinger:2018brx}.\footnote{
Here the symbol~$\tilde{A}$~is used to label the two-mass OMEs and has not the meaning as in
Eq.~(\ref{eq:til}).
The three-loop two-mass OMEs $(\Delta) \tilde{A}_{Qg}$ are in preparation \cite{TWOMAQG}.} The two-mass 
VFNS to three-loop order, in which also these quantities emerge, will be
presented in a forthcoming publication. 

The present paper provides the VFNS to three-loop order in the unpolarized and polarized single-mass case 
for the first time for phenomenological use for future simulations and inclusive data analyses. This comprises
a full set of precise and fast numerical representations of the contributing OMEs in $x$-space. 
Here $x$ denotes the Bjorken variable, which is identical to the momentum fraction $z$ of the respective 
parton in the infinite momentum frame, considering only twist-2 contributions.
Numerical illustrations for the matching relations for the different parton distributions are given. 
 
The paper is organized as follows. In Section~\ref{sec:2} we describe the basic formalism for the
single-mass variable flavor number scheme. The relation between the fixed and variable flavor number scheme
for the unpolarized structure functions $F_2(x,Q^2)$ are discussed in Section~\ref{sec:3}. Corresponding
relations apply to the polarized structure $g_1(x,Q^2)$ synonymously. Note that the asymptotic 
heavy-flavor corrections for charged current processes are given by partly different OMEs than 
those in the 
neutral current case. There are massless-massive single quark excitations, which are absent in the 
neutral current case. Consequently, the associated heavy-flavor corrections are different, 
cf.~Ref.~\cite{Buza:1997mg} and corrections given in Ref.~\cite{Blumlein:2014fqa}. In the present 
paper we 
will majorly consider the neutral current case. Numerical illustrations in the single-mass 
case are given in Section~\ref{sec:4} and the conclusions in Section~\ref{sec:5}. In the appendices 
we provide technical aspects as reference points for  the {\tt Fortran}-codes, {\tt OMEUNP3} 
and {\tt OMEPOL3} in the unpolarized and polarized cases from tree level to three-loop order in 
the Larin scheme \cite{Larin:1993tq}.\footnote{We note that in the calculation 
of  polarized anomalous dimensions 
and the massless and massive Wilson coefficients in deep-inelastic scattering for the QCD corrections 
only up to two Levi-Civita tensors contribute. Furthermore, the particle-polarizations are due to
{\it external} initial conditions in the photon case. 
For weak-boson exchange only the couplings of a single weak gauge
boson is considered, which is equivalent to the previous case.} 
Up to two-loop order we provide also the code for the polarized OMEs in the $\overline{\rm MS}$-scheme 
and for transversity, cf.~Refs.~\cite{Blumlein:2009rg,Ablinger:2014vwa}, {\tt OMEPOL2MS}.
In the three-loop case also a numerical code in {\tt C++} is provided. The 
associated {\tt Fortran} codes 
{\tt OMEUNP3} and {\tt OMEPOL3} are grid-based for fast performance. 
We also illustrate the constant part  of some unrenormalized single-mass OMEs 
$(\Delta) A_{ij}^{(3)}$, $(\Delta) a_{ij}^{(3)}$, 
for which this was not done before. These quantities contribute to the single-mass VFNS.
\section{The Basic Formalism}
\label{sec:2}

\vspace*{1mm}
\noindent
The transition formulae in the single-mass VFNS from three-loop order have been derived in Ref.~\cite{Buza:1996wv} 
and were corrected in Ref.~\cite{Bierenbaum:2009mv}. Here we consider first relations stemming from
neutral current structure functions with $f_i + f_{\bar{i}}$ quark content, implying non-singlet 
+-combinations.

We define the non-singlet and singlet parton distribution 
functions, $\Delta_{i}^{\rm NS,+}$ and $\Sigma$, by
\begin{eqnarray}
\Delta_{i}^{\rm NS,+}(x,Q^2,N_F) &=&  f_i(x,Q^2) + f_{\bar{i}}(x,Q^2) - \frac{1}{N_F} \Sigma(x,Q^2,N_F)
,~~~i \in \{1, \ldots, N_F\},
\\
\Sigma(x,Q^2,N_F) &=&  \sum_{i=1}^{N_F} \left[f_i(x,Q^2) + f_{\bar{i}}(x,Q^2)\right],
\end{eqnarray}
where $x$ denotes the longitudinal momentum fraction of the twist-2 parton distributions, $Q^2 = -q^2$ 
the virtuality of the space-like momentum transfer $q$, and $N_F$ the number of massless flavors.
We set $f_{i \pm \bar{i}}(x,Q^2) \equiv f_i(x,Q^2) \pm f_{\bar{i}}(x,Q^2)$. 

Sometimes it is convenient to work in Mellin $N$ space
since simpler relations are obtained there.
The Mellin transforms of a regular function, resp.
  +- and $\delta$-distributions, are given by 
\begin{eqnarray} 
  \Mvec[f(x)](N)     &=& \int_0^1 dx x^{N-1} f(x),\\
  \Mvec[[g(x)]_+](N) &=& \int_0^1 dx \left[x^{N-1}-1\right] g(x), \\ 
  \Mvec[[h(x) \delta(1-x)](N) &=& h(1),
\end{eqnarray} 
with $f(x), g(x), h(x) \in C[0,1]$. 
In Mellin space the transition formulae in the unpolarized case read\footnote{Note a series of 
typographical errors in Ref.~\cite{Buza:1996wv}, which were corrected in 
Ref.~\cite{Behring:2014eya}.}
\begin{eqnarray}
\label{match:1}
f_{i + \bar{i}}(N,Q^2,N_F+1) &=&  
  A_{qq,Q}^{\rm NS,+} \cdot f_{i + \bar{i}}(N,Q^2,N_F) 
+ \tilde{A}_{qq,Q}^{\rm PS} \cdot \Sigma(N,Q^2,N_F) 
\nonumber\\ &&
+ \tilde{A}_{qg,Q} \cdot G(N,Q^2,N_F),
\\
\Sigma(N,Q^2,N_F+1) &=&  
\Bigl[
  A_{qq,Q}^{\rm NS,+} 
+ A_{qq,Q}^{\rm PS} 
+ {A}_{Qq}^{\rm PS} \Bigr]   \cdot \Sigma(N,Q^2,N_F) 
  \nonumber\\  &&
+ \Bigl[
  A_{qg,Q} 
+ A_{Qg}  \Bigr]   \cdot G(N,Q^2,N_F),
\\
\label{match:2}
G(N,Q^2,N_F+1) &=&  
   A_{gq,Q} \cdot \Sigma(N,Q^2,N_F)
+  A_{gg,Q} \cdot G(N,Q^2,N_F).
\end{eqnarray}

The PDFs are no observables. By using either the FFNS or the VFNS, the structure 
functions
have to be the same in the limit of large virtualities. What happens changing from the 
FFNS to the VFNS is that the PDFs 
absorb the massive OMEs and the Wilson coefficients become massless ones. In course of 
this, heavy flavor PDFs are defined.

The massive OMEs are given by
\begin{eqnarray}
A_{ij} = A_{ij}\left(N, \frac{m^2_Q}{Q^2}\right) = \delta_{ij} + \sum_{k=1}^\infty a_s^k 
A_{ij}^{(k)},
\end{eqnarray} 
where $a_s = \alpha_s/(4\pi) = g^2_s/(4 \pi)^2$ denotes the strong coupling constant at the virtuality $Q^2$.
$G$ denotes the gluon distribution.
We use the notation
\begin{eqnarray} 
\label{eq:til}
\tilde{f}(N_F) \equiv \frac{f(N_F)}{N_F},~~~~~~~~~~~\hat{f}(N_F) \equiv f(N_F+1) - f(N_F).
\end{eqnarray} 
A few comments are in order. The factors $1/N_F$ in front of $A_{qq,Q}^{\rm PS}$
and $A_{qg,Q}$ in 
Eq.~(\ref{match:1})
account for the fact that the distributions are initially calculated not taking the charges
$e_f^2$ into account. Both in the singlet and 
pure-singlet cases the closed light-flavor loop carrying the operator insertion generates a factor 
of $N_F$. In the 
non-singlet case the external light quark line moves through the whole OME diagram on 
which the 
local operator and electromagnetic charge are sitting. Furthermore, one has to observe the explicit 
$N_F$-structure of the massless Wilson coefficients \cite{Vermaseren:2005qc,Blumlein:2022gpp} which 
we used to simplify the expressions.

The relations (\ref{match:1}--\ref{match:2}) apply to the transition from $N_F = 3$ to $N_F = 4$ flavor
PDFs through the non-vanishing OMEs $A_{ij}$, which depend on the quark mass of the 
heavy quark 
logarithmically via $\ln(m^2_Q/Q^2)$.
At tree level only the diagonal elements
    $A_{qq,Q}^{\rm NS}$ and $A_{gg,Q}$ do not vanish and evaluate to unity.
The first contributions are
those due to charm, $m_Q = m_c$. The single-mass bottom contributions are obtained in the same way 
and depend on  
$\ln(m_b^2/Q^2)$, leaving out the tree-level contributions. This is not yet the complete two-mass 
description, which also contains additional reducible and irreducible two-mass terms. 

Furthermore, the heavy-flavor distribution $f_{Q + \bar{Q}}$
\begin{eqnarray}
\label{match:3}
f_{Q +\bar{Q}} \equiv
f_{Q}(N,Q^2,N_F+1) 
+ f_{\bar{Q}}(N,Q^2,N_F+1) 
&=&  
   {A}_{Qq}^{\rm PS} \cdot \Sigma(N,Q^2,N_F)
+  A_{Qg} \cdot G(N,Q^2,N_F)\
\nonumber\\
\end{eqnarray}
is implied. The OMEs $A_{qg,Q}$ and $A_{qq,Q}^{\rm PS}$ only contribute from three-loop order.
At threshold, $Q^2 = m_Q^2$, the relation
\begin{eqnarray}
f_{Q}(N,Q^2,N_F+1) =  f_{\bar{Q}}(N,Q^2,N_F+1) 
\end{eqnarray}
holds. Between the thresholds at $Q^2 = m_c^2$ and $Q^2 = m_b^2$ the evolution of the parton densities 
is 
again massless. Due to non-singlet flavor combinations depending on the color factor $d_{abc}$, see
Refs.~\cite{Moch:2004pa,Blumlein:2021enk}, a difference emerges
in the now massless heavy-quark contributions as charm and at higher scales bottom.
For $N_F > 3$ also contributions of the kind contribute in the massless Wilson 
coefficients, see Refs.~\cite{Vermaseren:2005qc,Blumlein:2022gpp},
depending on the observable.  

There are three sum rules in the case of the unpolarized OMEs
\begin{eqnarray}
A_{qq,Q}^{\rm NS,-}(N=1) &=& 0,
\\
A_{qq,Q}^{\rm NS,+}(N=2) 
+ A_{qq,Q}^{\rm PS}(N=2) 
+ A_{Qq}^{\rm PS}(N=2) 
+ A_{gq,Q}(N=2)  &=& 1,
\\
A_{qg,Q}(N=2) 
+ A_{Qg}(N=2) 
+ A_{gg,Q}(N=2) &=& 1,
\end{eqnarray} 
due to fermion-number and energy-momentum conservation, here expressed in terms of moment-relations.

In  $x$-space the products in Mellin space turn into the following convolutions, 
cf.~Eqs.~(256--258) of Ref.~\cite{Bierenbaum:2022biv},
  \begin{eqnarray} 
\label{eq:con1}
  \left(\left[\frac{f}{1-x}\right]_+ \otimes g\right)(x) &=& \int_x^1 
  dz \frac{f(z)}{1-z} \left[\frac{1}{z} g\left(\frac{x}{z}\right) - g(x) \right]
- 
g(x) \int_0^x dz 
  \frac{f(z)}{1-z}, \\ 
\label{eq:con2}
(h \otimes g)(x) &=& \int_x^1 \frac{dz}{z} h(z) g\left(\frac{x}{z}\right), 
\\ 
\label{eq:con3}
(\delta(1-x) \otimes g)(x) &=& 
  g(x), 
\end{eqnarray} 
for the 
  +-distributions, regular functions, and the contributions containing $\delta(1-x)$. The Mellin 
convolutions
$\otimes$ are obtained from the general relation
\begin{eqnarray} 
[A \otimes B](x) = 
\int_0^1 dx_1 \int_0^1 dx_2 \delta(x - x_1 x_2) A(x_1) B(x_2).
\end{eqnarray} 

Since the mass ratio $m_c^2/m_b^2 \sim 1/10$ is not particularly small, one may consider to 
decouple charm and 
bottom together at a sufficiently high scale $Q^2$, cf.~Ref.~\cite{Ablinger:2017err}. In particular 
it is difficult 
to believe that charm can be considered massless at the mass scale of the bottom quark, $m_b$. Indeed it has been 
shown, e.g., for the unpolarized and polarized Bjorken sum-rule and the Gross-Llewellyn Smith sum 
rule in 
Ref.~\cite{Blumlein:2016xcy}, where to two-loop order the complete mass-dependence is known, that the massless
limits for charm and bottom are approached only at rather large scales $Q^2$. A numerical 
study to 
two-loop order
on the VFNS, based on Refs.~\cite{Laenen:1992zk,Laenen:1992xs,Riemersma:1994hv}, has been performed 
in Ref.~\cite{Alekhin:2009ni} 
with the same conclusions. This has also been shown in the pure-singlet case at two-loop order in 
Refs.~\cite{Blumlein:2019zux,Blumlein:2019qze}. Therefore, heavy-quark distributions 
cannot be introduced at $Q^2 = m^2_Q$, the heavy-flavor threshold used traditionally, as 
massless 
distributions.
Doing this still has to be considered rather as a {\it convention}. However, in all this 
the correct 
heavy-flavor representation
of the deep-inelastic structure functions has to be maintained. In particular, power 
corrections 
\cite{Buza:1995ie,Buza:1996xr,Blumlein:2016xcy,Blumlein:2019zux,Blumlein:2019qze} have to be
of negligible size in the kinematic regions considered.

The PDF matching, even in the later two-mass case, is performed independently of the 
electroweak couplings of the partons 
for the distributions $\Delta_i^{\rm NS,+}, \Sigma$ and $G$, unlike the case for the 
massive Wilson 
coefficients in 
deep-inelastic scattering. The relations are structurally the same as in 
Eqs.~(\ref{match:1}--\ref{match:2}) by 
just extending the respective 
  OMEs by their two-mass contributions $\tilde{A}_{qq,Q}^{\rm NS}, \tilde{A}_{Qq}^{\rm PS}, \tilde{A}_{Qg}, 
  \tilde{A}_{gq,Q}$ and $\tilde{A}_{gg,Q}$. Here $\tilde{A}_{Qg}$ and $\tilde{A}_{gg}$ start at 
two-loop order \cite{Blumlein:2018jfm} and the other OMEs at 
  three-loop order. At four-loop order there are also two-mass contributions for $\tilde{A}_{qq,Q}^{\rm PS}$ 
  and $\tilde{A}_{qg,Q}^{\rm PS}$. 

  Synonymous relations hold in the polarized case for the structure functions at longitudinal 
  polarization, cf.~Ref.~\cite{Blumlein:1996vs}, as, e.g., for the structure function $g_1(x,Q^2)$. 
Here we 
  work in the Larin scheme 
  \cite{Larin:1993tq}.\footnote{In previous papers the results in the non-singlet case were 
  often presented in the
  $\overline{\rm MS}$ scheme. The transition to the Larin scheme used in the present paper has been 
  given in Refs.~\cite{Schonwald:2019gmn,Blumlein:2021xlc}.} Accordingly, the evolution 
of the polarized parton 
  distributions is 
  performed in the Larin scheme, cf. Ref.~\cite{Blumlein:2024euz}. Corresponding transformations of the 
 anomalous 
  dimensions have been given in Refs.~\cite{Matiounine:1998re,Moch:2014sna,Blumlein:2021ryt}.

  Regarding the consideration of the evolution of parton densities as defined in the 
variable flavor
  number scheme, the following convention is used. The flavor thresholds are chosen to be the 
heavy-quark masses, either in the on-mass-shell scheme (OMS) 
  or the $\overline{\rm MS}$ scheme. Accordingly,
  one has to use the OMEs $A_{ij}$ with the heavy quark mass renormalized in these 
schemes\footnote{In the present paper we use the OMS scheme.
  The transition to the $\overline{\rm MS}$ scheme is easily obtained by using the relations between the 
  different mass definitions, cf., e.g., Ref.~\cite{Marquard:2016dcn} and references 
therein.} by expanding in the 
  coupling constant to the respective order.

  Furthermore, one wants to separate the quark and anti-quark distributions within this framework. This
  is not possible referring only to the structure function $F_2(x,Q^2)$ 
  in the unpolarized case or
  $g_1(x,Q^2)$ in the polarized case of either proton or deuteron targets only.
  Rather one considers additionally other flavor non-singlet combinations from non-singlet structure 
  functions, e.g. the combination of structure functions $xF_3^{W^+-W^-}(x,Q^2)$. Corresponding 
  relations are given in Ref.~\cite{Behring:2015roa}. Here, the heavy-quark contributions 
depend on 
  $A_{qq,Q}^{\rm NS}$
  analytically continued from the odd moments.\footnote{For the different nucleon targets also the
  Cabibbo-Kobayashi-Maskawa matrix elements contribute.} For the corresponding contributions in the 
  polarized case for the (target-projected) non-singlet contribution see 
Ref.~\cite{Behring:2015zaa}.   
  One obtains
\begin{eqnarray}
f_i(N,Q^2,N_F+1) - f_{\bar{i}}(N,Q^2,N_F+1) = A_{qq,Q}^{\rm NS,-}
[f_i(N,Q^2,N_F) - {f}_{\bar{i}}(x,Q^2,N_F)] 
\end{eqnarray}
in this case.
\section{\boldmath $F_2(x,Q^2)$ and $g_1(x,Q^2)$ in the Fixed and Variable Flavor Number Schemes}
\label{sec:3}

\vspace*{1mm}
\noindent

In the following we discuss the relations for the unpolarized structure function $F_2(x,Q^2)$ by expressing this 
observable either in the fixed or the variable flavor number scheme. Corresponding relations hold for the 
polarized structure function $g_1(x,Q^2)$ synonymously. In the fixed flavor number scheme the structure functions
obey the representation
\begin{eqnarray} 
\label{eq:F2tot}
F_2^{\rm total}(x,Q^2) = F_2^{\rm light}(x,Q^2) + F_2^{\rm Q}(x,Q^2),
\end{eqnarray} 
where the massless contributions $F_2^{\rm light}(x,Q^2)$ are given in Eqs.~(87, 92) of 
Ref.~\cite{Blumlein:2022gpp} for $N_F$ massless flavors, see also 
Ref.~\cite{Vermaseren:2005qc}. The single-mass heavy-flavor contributions 
$F_2^{\rm Q}(x,Q^2)$ are given in Ref.~\cite{Bierenbaum:2009mv}, Eqs.~(2.1) and 
(2.11--2.15).

Observables, like the deep-inelastic structure functions $F(x,Q^2)$, obey the 
renormalization group equation (RGE), cf.~Ref.~\cite{Blumlein:2000wh},
\begin{eqnarray} 
\label{eq:RGE2}
\Da F(x,Q^2) = 0
\end{eqnarray} 
with the RGE operator
\begin{eqnarray}
\label{eq:Da} 
\Da  = \mu \frac{\partial}{\partial \mu} 
+ \beta(g,N_F) \frac{\partial}{\partial g} 
\end{eqnarray} 
and $g=g(\mu^2, N_F) = 4 \pi \sqrt{a_s}$. The heavy-quark masses are fixed 
parameters in the on-shell scheme. In the present approach they only emerge 
correlated with the scale $\mu^2$. 
The evolution equations for the massless off-shell OMEs $\Gamma_{ij}$ at space-like 
momentum square $p^2$, cf.~Ref.~\cite{Zijlstra:1992qd}, form the twist-2 parton 
densities. 
Their evolution equations and the ones of the massless
on-shell Wilson coefficients read
\begin{eqnarray} 
\Da \Gamma_{ij}(\mu^2/p^2) = - \gamma_{ik} \Gamma_{kj}(\mu^2/p^2),~~~~
\Da C_{i,l}(Q^2/\mu^2) = \gamma_{lm} C_{i,m}(Q^2/\mu^2),~~~~
\end{eqnarray} 
using the Einstein summation convention. Here $\gamma_{ik} = \gamma_{ik}(a_s)$ are the anomalous dimensions.
The massless PDFs obey
\begin{eqnarray} 
\Da \Delta_k^{\rm NS,+}(\mu^2) &=& - \gamma_{qq}^{\rm NS}~\Delta_k^{\rm NS,+}(\mu^2), 
\\
\Da \Sigma(\mu^2)   &=& - [\gamma_{qq}^{\rm S}~\Sigma(\mu^2) + \gamma_{qg}^{\rm 
S}~G(\mu^2)], 
\\
\Da G(\mu^2)        &=& - [\gamma_{gq}^{\rm S}~\Sigma(\mu^2) + \gamma_{gg}^{\rm 
S}~G(\mu^2)]. 
\end{eqnarray} 
The action of the RGE operator $\Da$ can be transformed into the derivative for the 
variable
\begin{eqnarray} 
t = - 2 \ln\left(\frac{\mu_0^2}{\mu^2}\right),
\end{eqnarray} 
through which the evolution equations for the massless Wilson coefficients and parton 
densities read  
\begin{eqnarray} 
\frac{d}{dt} C_i(t) =    \gamma_{ij}(t) C_j(t),~~~~~ 
\frac{d}{dt} f_i(t) =  - \gamma_{ij}(t) f_j(t). 
\end{eqnarray} 
Here $\mu_0$ denotes a mass scale and
\begin{eqnarray} 
\frac{d a_s(t)}{dt} = - 2 \sum_{k=0}^\infty \beta_k a_s^k(t). 
\end{eqnarray} 
The anomalous dimensions are given by
\begin{eqnarray} 
\gamma_{ij}(t) =\sum_{k=0}^\infty \gamma_{ij}^{(k)} a_s^l(t).
\end{eqnarray} 
Equation~(\ref{eq:RGE2}) leads to 
\begin{eqnarray} 
\label{eq:RGE3}
\frac{d}{dt} F(t) = 0.
\end{eqnarray} 

The expression for $F_2(x,Q^2)$ in the variable flavor number scheme reads
\begin{eqnarray} 
\label{eq:V1}
F_2^{\rm VFNS}(x,Q^2) = F_2^{\rm VFNS, e_k^2}(x,Q^2) + F_2^{\rm VFNS, e_Q^2}(x,Q^2),
\end{eqnarray} 
with 
\begin{eqnarray} 
\label{eq:massl1}
F_2^{\rm VFNS, e_k^2}(x,Q^2) &=& \sum_{k=1}^{N_F} e_k^2 \Biggl[
f_{k+\bar{k}}(x,Q^2,N_F+1) \otimes
\Ca_2^{\rm NS}\left(x,\frac{Q^2}{\mu^2},N_F+1\right)
\nonumber\\ &&
+ \Sigma(x,\mu^2,N_F+1) \otimes
\tilde{\Ca}_{2,q}^{\rm PS}\left(x,\frac{Q^2}{\mu^2},N_F+1\right)
\nonumber\\ &&
+ G(x,\mu^2,N_F+1) \otimes
\tilde{\Ca}_{2,g}^{\rm S}\left(x,\frac{Q^2}{\mu^2},N_F+1\right)\Biggr],
\\
F_2^{\rm VFNS, e_Q^2}(x,Q^2) &=& e_Q^2 \Biggl[
f_{Q+\bar{Q}}(x,Q^2,N_F+1) \otimes
\Ca_2^{\rm NS}\left(x,\frac{Q^2}{\mu^2},N_F+1\right)
\nonumber\\ &&
+ \Sigma(x,\mu^2,N_F+1) \otimes
\tilde{\Ca}_{2,q}^{\rm PS}\left(x,\frac{Q^2}{\mu^2},N_F+1\right)
\nonumber\\ &&
+ G(x,\mu^2,N_F+1) \otimes
\tilde{\Ca}_{2,g}^{\rm S}\left(x,\frac{Q^2}{\mu^2},N_F+1\right)\Biggr],
\end{eqnarray} 
where $\Ca_{2,i}$ denote the massless Wilson coefficients. Using the above relations,
it is easy to show that the two terms in the l.h.s.\ of Eq.~(\ref{eq:V1})
are individually obeying Eq.~(\ref{eq:RGE2}).
However, the evolution of the parton distribution functions now depends on the massive 
OMEs, $A_{ij}$, exhibiting a more involved scale dependence, cf.\ Ref.~\cite{Buza:1996wv} 
and Eqs.~(\ref{match:1}--\ref{match:2}, \ref{match:3}).  
$F_2^{\rm VFNS, e_k^2}(x,Q^2)$ 
also contains the complete massless contribution for the  $N_F$ light flavors, 
(\ref{eq:massl1}), which has to be subtracted to obtain the corresponding heavy-flavor 
term. 

We have shown by an explicit calculation that for $Q^2 \gg m_Q^2$, 
\begin{eqnarray} 
\delta F_2(x,Q^2) = 
F_2^{\rm total}(x,Q^2) - F_2^{\rm VFNS}(x,Q^2) = O(a_s^4),
\end{eqnarray} 
holds by referring to the massive OMEs and massless Wilson coefficients to 
three-loop order.
In the representations given in Ref.~\cite{Bierenbaum:2009mv,Behring:2014eya},
both for the structure functions as well as for the relations in the VFNS, we did not write quantities
which vanish in QCD in the respective order. One example is $\hat{C}_{2,q}^{(1),\rm NS} = 0$, which is implied 
by the concrete $N_F$ dependence.
Corresponding representations hold for the structure function $g_1(x,Q^2)$. 

In any other representation or scheme used one has to provide the observables, the deep-inelastic 
structure functions and others, including the heavy-flavor contributions correctly, i.e., 
up to 
$O(a_s^3)$, 
in the present case. 
\section{Numerical Results}
\label{sec:4}

\vspace*{1mm}
\noindent
To illustrate the single-mass heavy-flavor contributions to the parton distribution 
functions  
in 
the VFNS for charm and bottom in relation to those for light flavors, we use $x$-space 
parameterizations of the kind
\begin{equation}
\label{eq:DIST}
x f_i(x) = N_i x^{a_i} (1-x)^{b_i} (1 + c_i x^{d_i})
\end{equation}
for each value of $Q^2$ of the non-singlet, singlet and gluon distributions, 
similar to~Ref.~\cite{Blumlein:2006mh}.\footnote{Note a typographical error in 
Eqs.~(54,55) 
in Ref.~\cite{Blumlein:2006mh}.} The parameters for Eq.~(\ref{eq:DIST}) are given 
in Table~\ref{TAB1} in Appendix~\ref{sec:A} represent an effective parameterization in 
the unpolarized case. The distributions are positive in the whole $x$-range.
\footnote{We thank H.~B\"ottcher for providing this parameterization.}
\begin{figure}[H]
\centering
\includegraphics[width=0.49\textwidth]{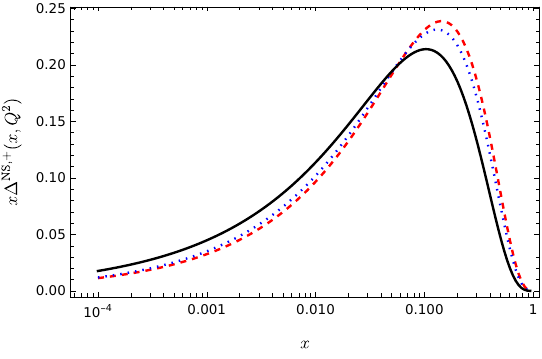}
\caption{\sf The distribution $x\Delta^{\rm NS,+}(x,Q^2)$ for
$Q^2 =    30~\GeV^2$ (dotted line),
$Q^2 =   100~\GeV^2$ (dashed line),
$Q^2 = 10000~\GeV^2$ (full  line) in the unpolarized case.
\label{fig:1}}
\end{figure}

\begin{figure}[H]
\centering
\includegraphics[width=0.49\textwidth]{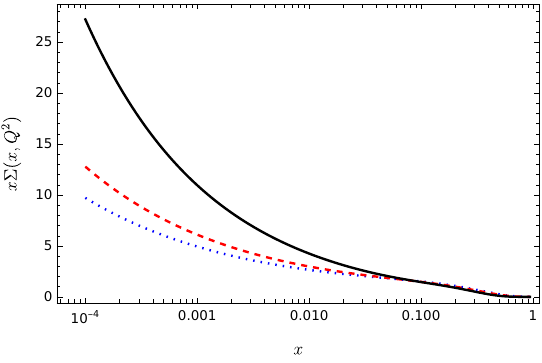}
\includegraphics[width=0.49\textwidth]{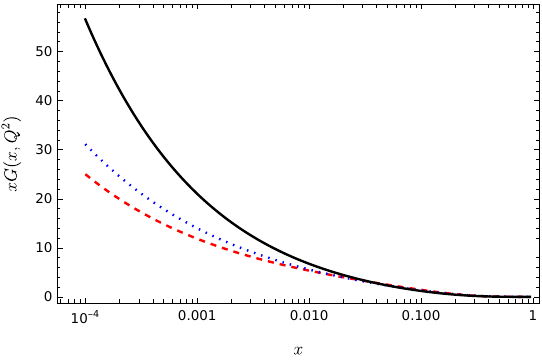}
\caption{\sf Left panel: The distribution $x\Sigma(x,Q^2)$ for
$Q^2 =    30~\GeV^2$ (dotted line),
$Q^2 =   100~\GeV^2$ (dashed line),
$Q^2 = 10000~\GeV^2$ (full  line) in the unpolarized case.
Right panel: the same for the distribution $xG(x,Q^2)$.
\label{fig:2}}
\end{figure}

To establish a baseline, we first show the massless distributions before illustrating the 
impact of the heavy flavor contributions.
In Figures~\ref{fig:1} and \ref{fig:2} we show the $x$ and $Q^2$  dependence of the 
unpolarized parton distributions for the massless partons. While the non-singlet distribution falls 
towards the region of small values of $x$ those of the singlet and gluon momentum distributions are 
rising.  The singlet and gluon distributions are related by four-momentum conservation,  
\begin{equation}
\int_0^1 dx x[\Sigma(x,Q^2) + G(x,Q^2)] = 1.
\end{equation}

In the polarized case to three-loop order, we use the distributions in the 
Larin scheme of 
Ref.~\cite{Blumlein:2024euz}.\footnote{The non-singlet OMEs are also available in the 
$\overline{\rm MS}$ scheme.} Also the polarized three-loop massless singlet 
Wilson coefficients are only available in the Larin scheme, cf.\ Ref.~\cite{Blumlein:2022gpp}.
It is useful to express the individual polarized parton densities for three light flavors 
in terms of  the distributions $x\Delta_3, x\Delta_8, x\Delta \Sigma$,  
and $x\Delta G$. From 
the relations
\begin{eqnarray}
\Delta_3 &=& (\Delta u + \Delta \bar{u}) - (\Delta d + \Delta \bar{d}), \\ 
\Delta_8 &=& (\Delta u + \Delta \bar{u}) + (\Delta d + \Delta \bar{d}) - 2(\Delta s + \Delta \bar{s}),\\ 
\Delta \Sigma &=& \Delta u + \Delta \bar{u} + \Delta d + \Delta \bar{d} +
\Delta s + \Delta \bar{s},
\end{eqnarray}
one obtains
\begin{eqnarray}
\Delta u + \Delta \bar{u} &=& \frac{1}{6} \left[\Delta_8 + 2 \Delta \Sigma + 3 \Delta_3\right],
\\
\Delta d + \Delta \bar{d} &=& \frac{1}{6} \left[\Delta_8 + 2 \Delta \Sigma - 3 \Delta_3\right],
\\
\Delta s + \Delta \bar{s} &=& \frac{1}{3}  \left[\Delta \Sigma - \Delta_8\right].
\end{eqnarray}

\begin{figure}[H]
\centering
\includegraphics[width=0.49\textwidth]{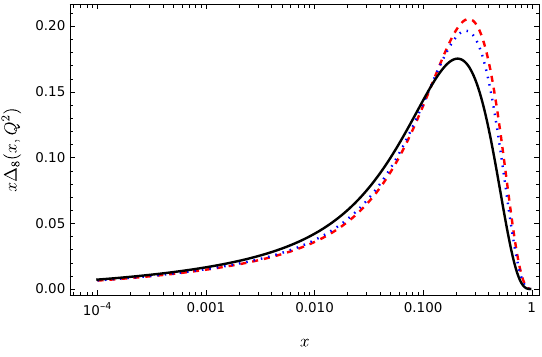}
\caption{\sf The distribution $x\Delta_8(x,Q^2)$ for
$Q^2 =    30~\GeV^2$ (dotted line),
$Q^2 =   100~\GeV^2$ (dashed line),
$Q^2 = 10000~\GeV^2$ (full  line) in the polarized case in the Larin scheme 
\cite{Blumlein:2024euz}.
\label{fig:3}}
\end{figure}

\vspace*{-10mm}
\begin{figure}[H]
\centering
\includegraphics[width=0.49\textwidth]{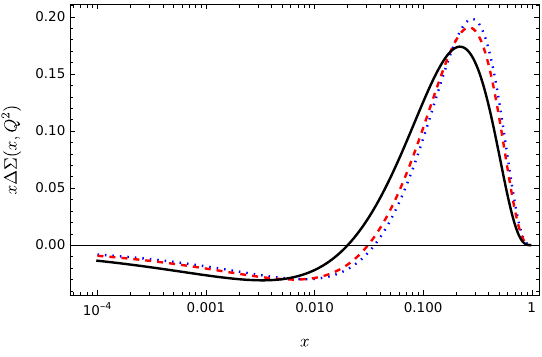}
\includegraphics[width=0.49\textwidth]{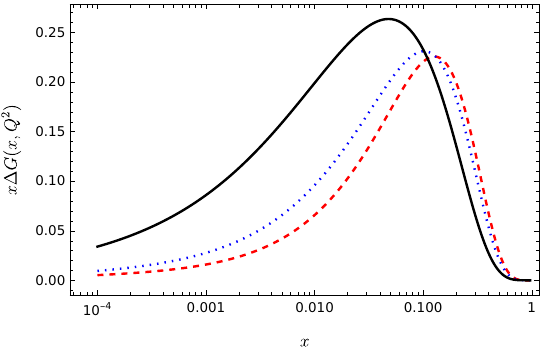}
\caption{\sf Left panel: the distribution $x\Delta\Sigma(x,Q^2)$ for
$Q^2 =    30~\GeV^2$ (dotted line),
$Q^2 =   100~\GeV^2$ (dashed line),
$Q^2 = 10000~\GeV^2$ (full  line) in the polarized case in the Larin scheme 
\cite{Blumlein:2024euz}.
Right panel: the same for the distribution  $x\Delta G(x,Q^2)$.
\label{fig:4}}
\end{figure}

Assuming a $SU(3)$ symmetric sea one has
\begin{eqnarray}
\Delta \bar{q} &=&  \Delta s = \Delta \bar{s} = \Delta \bar{u} = \Delta \bar{d}, 
\end{eqnarray}
which allows to form the combinations $\Delta f_i - \Delta f_{\bar{i}}$,
\begin{eqnarray}
\Delta u - \Delta \bar {u} &=& \frac{1}{2}\left[\Delta_8 + \Delta_3\right],\\
\Delta d - \Delta \bar {d} &=& \frac{1}{2}\left[\Delta_8 - \Delta_3\right],\\
\Delta s - \Delta \bar {s} &=& 0.
\end{eqnarray}
\vspace*{-7mm}
\begin{figure}[H]
\centering
\includegraphics[width=0.49\textwidth]{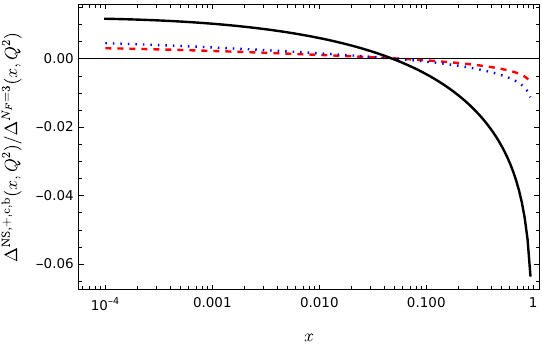}
\caption{\sf The ratio $\Delta^{\rm c,b}(x,Q^2)/\Delta^{\rm N_F =3}(x,Q^2)$ as a function of 
$x$, 
referring to the matching scales  
$Q^2 =    30~\GeV^2$ (dotted line),
$Q^2 =   100~\GeV^2$ (dashed line),
$Q^2 = 10000~\GeV^2$ (full  line) in the unpolarized case.
\label{fig:5}}
\end{figure}

The distribution functions for the massless partons are illustrated in 
Figures~\ref{fig:3} and \ref{fig:4}.
In the non-singlet case we will refer to $\Delta_8(x,Q^2)$, cf.~Ref.~\cite{Blumlein:2021lmf}, which is the 
only non-singlet combination in the case of the structure function $g_1^d(x,Q^2)$.
Unlike in the unpolarized case, the PDFs $x\Delta_8, x\Delta \Sigma$ and $x\Delta G$ in the polarized 
case are of comparable size.

We consider ratios relating the PDFs in the VFNS to the respective
$N_F = 3$ distributions at a number of scales $Q^2$. The 
ratios depend to a lesser extent on the 
details of the respective input distributions than in the case of absolute predictions.
The massive OMEs have been calculated in the on-mass-shell scheme for the heavy quark masses. 
We use the following values for the on-shell quark masses
\cite{Alekhin:2012vu,Agashe:2014kda}, 
\begin{equation}
m_c = 1.59~\GeV,~~~~~~~m_b = 4.78~\GeV.
\end{equation}

The numerical integrals are evaluated with {\tt DAIND} \cite{AIND}. In part the numerical 
representations \cite{Gehrmann:2001pz,Ablinger:2018sat} of the
harmonic polylogarithms \cite{Remiddi:1999ew}  are used. In the case of functions depending
on even higher alphabets \cite{Ablinger:2017bjx,Blumlein:2018aeq,Abreu:2019fgk}, we use precise numerical 
expansions in the interval $x \in [0,1]$, 
cf.~Ref.~\cite{Behring:2023rlq,Fael:2022miw}, 
which
also imply the occurrence of multiple zeta values up to weight {\sf w = 5}~\cite{Blumlein:2009cf}.
Finally we accelerated the code performance by local series expansions and by using precise grid
interpolation being finite in $x \in [10^{-6},1]$ and analytic treatment of the regions 
near to $x=0$ and 
$x=1$. 
For some subroutines we used code optimization \cite{Ruijl:2017dtg}.
We illustrate the distributions obtained in the VFNS, matching charm and bottom
quark single-mass effects simultaneously. This neglects effects from the
genuine two-mass OMEs which will be dealt with in a forthcoming publication.
For the distributions $D \in \{\Delta_i^{\text{NS},+}, \Sigma, G, \Delta_8,
\Delta\Sigma, \Delta G\}$ we define the combined charm and bottom contributions
\begin{align}
  D^{c,b}(x,Q^2) = D^c(x,Q^2) + D^b(x,Q^2) - 2 D^{N_F=3}(x,Q^2) \,,
\end{align}
\begin{figure}[H]
\centering
\includegraphics[width=0.49\textwidth]{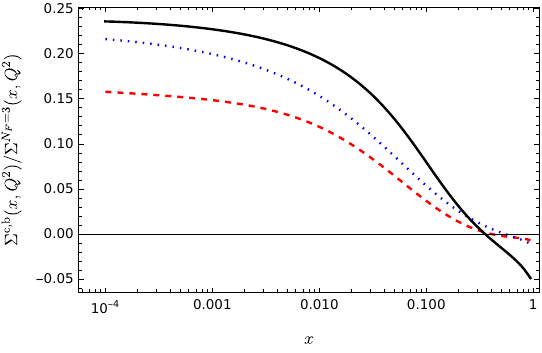}
\includegraphics[width=0.49\textwidth]{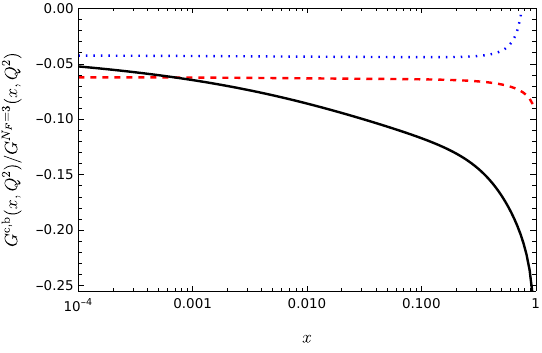}
\caption{\sf Left panel: The ratio $\Sigma^{\rm c,b}(x,Q^2)/\Sigma^{N_F =3}(x,Q^2)$ as a 
function of $x$, 
referring 
to the matching scales  
$Q^2 =    30~\GeV^2$ (dotted line),
$Q^2 =   100~\GeV^2$ (dashed line),
$Q^2 = 10000~\GeV^2$ (full  line) in the unpolarized case.
Right panel: the same for $G^{\rm c,b}(x,Q^2)/G^{N_F =3}(x,Q^2)$.
\label{fig:6}}
\end{figure}
\begin{figure}[H]
\centering
\includegraphics[width=0.49\textwidth]{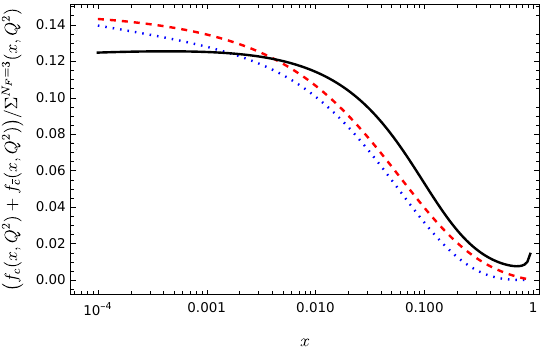}
\includegraphics[width=0.49\textwidth]{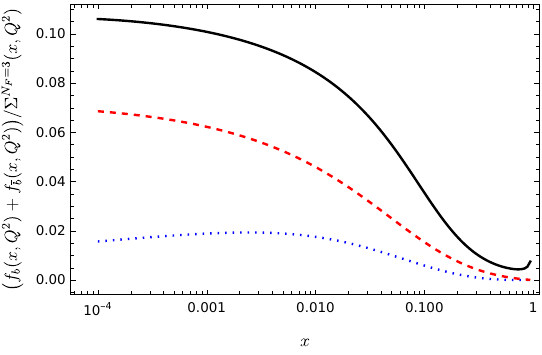}
\caption{\sf Left panel: the ratio $(f_c(x,Q^2) + f_{\bar{c}}(x,Q^2))/\Sigma^{N_F=3}(x,Q^2)$ as a function 
of $x$ referring, to the matching scales  
$Q^2 =    30~\GeV^2$ (dotted line),
$Q^2 =   100~\GeV^2$ (dashed line),
$Q^2 = 10000~\GeV^2$ (full  line) in the unpolarized case.
Right panel: the same for the ratio $(f_b(x,Q^2) + f_{\bar{b}}(x,Q^2))/\Sigma^{N_F=3}(x,Q^2)$. 
\label{fig:7}}
\end{figure}

\noindent
where $D^c(x,Q^2)$ and $D^b(x,Q^2)$ are the charm and bottom distributions
obtained from the matching relations according to Eqs. (7-9) at NNLO. Moreover,
$D^{N_F=3}(x, Q^2)$ are the corresponding three-flavor distributions that are
shown in Figures 1-4. Finally, we also consider the $c$- and $b$-quark
distributions obtained from Eq. (12).

The matching is done at the scale $Q^2$ and no further evolution is applied.
We show illustrations for $Q^2=30\,\mathrm{GeV}^2$, $100\,\mathrm{GeV}^2$ and
$10000\,\mathrm{GeV}^2$. We use the values 
$\alpha_s(30~\GeV^2) = 0.1972,~ 
\alpha_s(100~\GeV^2) = 0.1706,~
\alpha_s(10000~\GeV^2) = 0.1131$
for the strong coupling constant, consistent with the value 
$\alpha_s(M_Z^2) = 0.1147$.
The quark contributions will be normalized to the massless singlet 
distribution
and the gluon contributions to the massless gluon distribution at $N_F = 
3$.

\vspace*{-4mm}
\begin{figure}[H]
\centering
\includegraphics[width=0.49\textwidth]{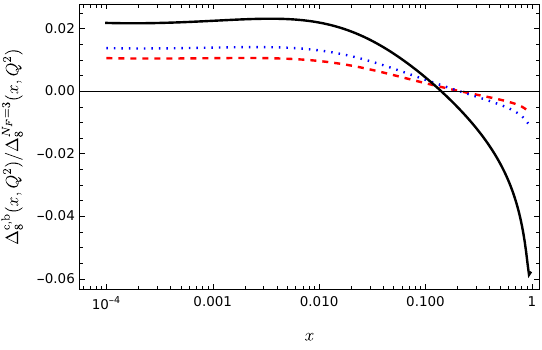}
\caption{\sf The ratio $\Delta_8^{\rm c,b}(x,Q^2)/\Delta_8^{\rm N_F =3}(x,Q^2)$ as a function 
of $x$, 
referring to the 
matching scales  
$Q^2 =    30~\GeV^2$ (dotted line),
$Q^2 =   100~\GeV^2$ (dashed line),
$Q^2 = 10000~\GeV^2$ (full  line) in the polarized case.
\label{fig:8}}
\end{figure}
\begin{figure}[H]
\centering
\includegraphics[width=0.49\textwidth]{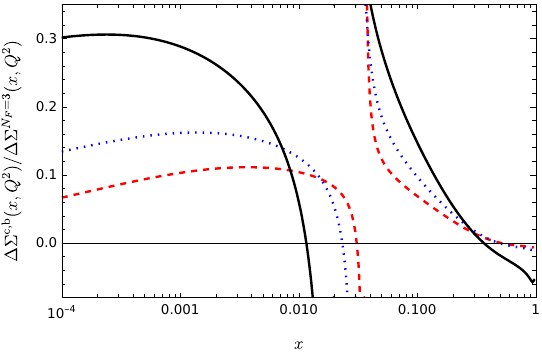}
\includegraphics[width=0.49\textwidth]{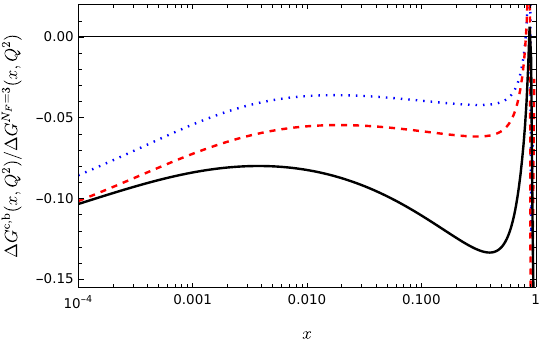}
\caption{\sf Left panel: the ratio $\Delta \Sigma^{\rm c,b}(x,Q^2)/\Delta \Sigma^{N_F =3}(x,Q^2)$ as a 
function of $x$, at  the matching scales  
$Q^2 =    30~\GeV^2$ (dotted line),
$Q^2 =   100~\GeV^2$ (dashed line),
$Q^2 = 10000~\GeV^2$ (full  line) in the polarized case.
Right panel: the same for the ratio $\Delta G^{\rm c,b}(x,Q^2)/\Delta G^{N_F =3}(x,Q^2)$.
\label{fig:9}}
\end{figure}

In Figure~\ref{fig:5} we illustrate the ratio $\Delta^{\rm c,b}(x,Q^2)/\Delta^{N_F =3}(x,Q^2)$ 
for different matching scales. The distribution $\Delta^{N_F =3}(x,Q^2)$ is a combination
of PDFs such that the singlet and gluon contributions to Eq.~(\ref{match:1}) cancel.
In the small-$x$ region the corrections range from 0.5\% to 
1.5\% and from $-1\%$ to $-6\%$ at 
large $x$.

Figure~\ref{fig:6} shows the analogous quantity for the singlet and the gluon 
distributions. $\Sigma^{c,b}$ amounts from 15\% to 25\% of $\Sigma^{\rm N_F=3}$ 
at $x=10^{-4}$ from $Q^2 = 30$ to $10^4~\GeV^2$ and falls toward large values of $x$. 
$G^{c,b}$ takes negative values and varies between $-5\%$ and $~\sim -15\%$ of   
$G^{\rm N_F=3}$. In the large $x$ region ratios of $O(-25\%)$ are obtained
for $ Q^2 = 10^4~\GeV^2$.

In Figure~\ref{fig:7} the ratio of the charm and bottom quark 
distributions $f_Q^{N_F=5}(x,Q^2)
+ f_{\bar{Q}}^{N_F=5}(x,Q^2)$, $Q = c,b$, to the singlet distribution $\Sigma^{N_F 
  =3}(x,Q^2)$ are shown. The effects at a given scale $Q^2$ are larger in the case of 
  charm than for bottom, which is due to the logarithmic contributions. The ratio grows towards
  small values of $x$.

Now we turn to the changes of the polarized distributions in the VFNS. In 
Figure~\ref{fig:8} the heavy-quark corrections to the non-singlet distribution 
$\Delta_8(x,Q^2)$ are shown. While at $Q^2 = 30~\GeV^2$ the ratio lays in the region 
$O(\pm 1\%)$ and grows for 
$Q^2 = 10000~\GeV^2$ to $\sim 2\%$ in the small-$x$ region and turns negative in the large-$x$
region becoming of the order of $-6\%$. This behavior is similar to that in the 
unpolarized case.

\begin{figure}[H]
\centering
\includegraphics[width=0.49\textwidth]{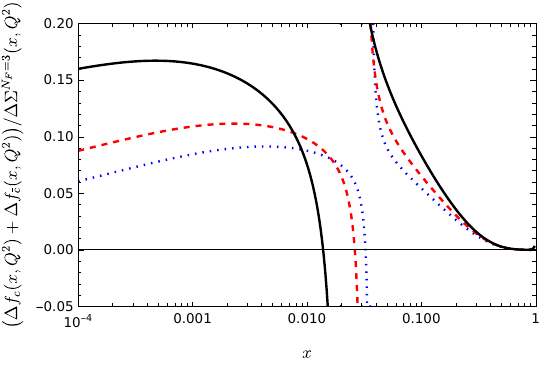}
\includegraphics[width=0.49\textwidth]{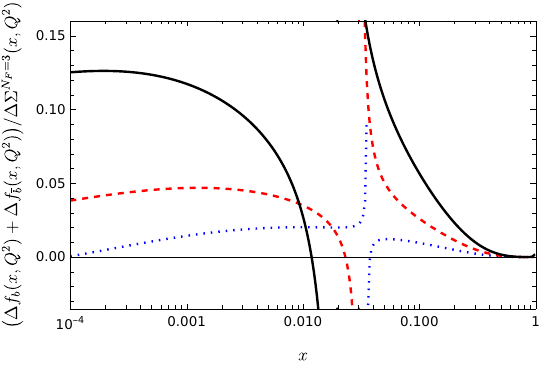}
\caption{\sf Left panel: the ratio $(\Delta f_c(x,Q^2) + \Delta f_{\bar{c}}(x,Q^2))/\Delta \Sigma^{N_F=3}(x,Q^2)$ as 
a function of $x$ referring, to the matching scales
$Q^2 =    30~\GeV^2$ (dotted line),
$Q^2 =   100~\GeV^2$ (dashed line),
$Q^2 = 10000~\GeV^2$ (full  line) in the unpolarized case.
Right panel: the same for the ratio $(\Delta f_b(x,Q^2) + \Delta f_{\bar{b}}(x,Q^2))/\Sigma^{N_F=3}(x,Q^2)$.
\label{fig:10}}
\end{figure}

The ratios of the heavy-flavor contributions to $\Delta \Sigma^{c,b}(x,Q^2)$ and $\Delta 
G^{c,b}(x,Q^2)$ to the respective light-flavor contributions are shown in Figure 
\ref{fig:9}.
In the small-$x$ region the ratio for $\Delta \Sigma(x,Q^2)$ grows from 0.1 to 0.3 from $Q^2 = 
30~\GeV^2$
to $Q^2 = 10^4~\GeV^2$. Around $x \sim 0.02$ the massless contributions change from negative to positive 
values. For larger values of $x$ the ratio drops and takes small negative values at very 
large $x$.
The ratios for $\Delta G(x,Q^2)$ start around $-0.1$ at small $x$ and rise to zero in the 
large-$x$ 
regions. At large $Q^2$, values of $\sim -0.15$ are reached.

The heavy-quark distributions $\Delta f_Q + \Delta f_{\bar{Q}}$ normalized to the 
polarized singlet distributions $\Delta \Sigma^{N_F =3}$ are shown in 
Figure~\ref{fig:10}.
The charm distributions are larger than those of bottom. At small $x$ and $Q^2 = 
10000~\GeV^2$ charm amounts to 16\% of the singlet distribution. The singularity
is due to the zero in $\Delta \Sigma^{N_F =3}$. The distributions grow with growing 
values of $Q^2$ and are also non-negligible in the polarized case. 

\section{Conclusions}
\label{sec:5}

\vspace*{1mm}
\noindent
We have presented the three-loop matching relations of the single-mass twist-2 VFNS in the unpolarized and 
polarized case for the first time, with the corresponding codes for phenomenological and data analysis. 
Seven single-mass OMEs contribute, supplemented by one more non-singlet 
OME to match also the combination $f_i - f_{\bar{i}}$. These OMEs represent also the 
massive Wilson coefficients in the asymptotic region $Q^2 \gg m^2_Q$, being the region in
which the 
matchings in the VFNS are derived. We have shown that for the inclusive structure 
functions $F_2(x,Q^2)$ 
and $g_1(x,Q^2)$ the representation in the fixed flavor number scheme and 
the variable flavor number scheme 
agree up to $O(a_s^3)$ in the coupling constant.
As has been shown in 
Ref.~\cite{Blumlein:2016xcy} already, rather high scales $Q^2$ are required to let 
the heavy-quark 
contributions to become light. Therefore, heavy-quark distributions cannot be introduced at $Q^2 
= 
m^2_Q$, the respective heavy-flavor threshold,
in a phenomenological sense, but rather constitute  a convention. We recall that the 
logarithms 
$\ln(m^2_Q/Q^2)$ are phase space 
logarithms, like in QED \cite{Beenakker:1989km,Blumlein:1989gk,Kripfganz:1990vm}, and not 
just collinear 
logarithms with a subsidiary scale $\mu = m_Q$, like in massless approaches 
\cite{Politzer:1974fr,Buras:1979yt,Reya:1979zk,Blumlein:2012bf,Blumlein:2023aso}. From the four 
different sources 
contributing to the renormalization of the massive OMEs, see Ref.~\cite{Bierenbaum:2009mv}, Section~3, 
only one, the collinear singularities, can be  absorbed into the scaling violations of the massless PDFs, 
due to mass-factorization in the $\overline{\rm MS}$-scheme, in which we work. 
The other logarithmic 
contributions $\ln(Q^2/m_c^2)$ and $\ln(Q^2/m_b^2)$ are no collinear 
logarithms, since they do, in general, not occur together with anomalous dimensions\footnote{
An exception is $(\Delta) A_{Qg}^{(1)}$, were $(\Delta) a_{Qg}^{(1)}$ vanishes, 
cf.~\cite{Buza:1995ie,Bierenbaum:2007qe,Buza:1996xr,Bierenbaum:2022biv}.} only,
like the collinear mass singularities. Here also finite expansion pieces of the massive OMEs, 
$(\Delta) 
a_{ij}$, contribute. In the present paper we do not resum these logarithms, like it has 
been done in Refs.~\cite{Cacciari:1998it,Forte:2010ta}, leaving a corresponding study for later work.

We have presented a series of numerical results on the changes of the non-singlet, singlet and gluon 
distribution functions by going from $N_F \rightarrow N_F + 2$, maintaining the 
single-mass 
corrections in the VFNS in the unpolarized and polarized cases. Here we performed the 
PDF matching at $Q^2 
= 30, 100$ and $10000~\GeV^2$. The corresponding corrections to the distributions $\Delta, \Sigma$ and $G$, 
and, similarly, to the corresponding polarized distributions, are large. Furthermore, the corresponding 
heavy-quark parton distribution functions have been derived in this context. 

The VFNS refers to only universal heavy-flavor corrections. There are process-dependent 
heavy-flavor 
contributions also contained in the Wilson coefficients, including power corrections, coming from phase-space 
integrals \cite{Buza:1995ie,Buza:1996xr,
Blumlein:2019zux,Blumlein:2016xcy,Blumlein:2019qze}, which have to be considered by assembling 
the PDFs of 
the VFNS framework and the Wilson coefficients to observables, as e.g. 
deep-inelastic structure functions. At large enough scales $Q^2$ these corrections become
negligible.

The {\tt Fortran} and {\tt C++}-codes providing the numerical representation of the matching coefficients
are attached to the arXiv-version of the paper as ancillary files. The license 
conditions are given in these codes.

\appendix
\section{Test values}
\label{sec:A}

\vspace*{1mm}
\noindent
The parameters for the different PDFs in $x$-space in the unpolarized case, Eq.~(\ref{eq:DIST}),
are given in Table~\ref{TAB1}.

In the following we provide a series of 
benchmark results to facilitate checks of the  implementation.
In Tables~\ref{TAB2}--\ref{TAB5} we present numerical results for the
OMEs in $x$-space for the unpolarized and polarized cases up to
three loops for $N_F=3$, respectively. 
In  Table~\ref{TAB2} we 
list $(\Delta) A_{ij}$ to three loops at $N_F = 3$, 
$\ln(m^2_Q/\mu^2) = 10$, $a_s = 1/10$  and $x = 1/10$, which are 
arbitrary numerical test values.  
In Table~\ref{TAB3} we list corresponding values for the polarized massive OMEs in the 
$\overline{\rm MS}$ scheme to two-loop order.

Finally, we list in Tables~\ref{TAB4} and \ref{TAB5} the moments $N = 2,4,6$ in the unpolarized 
case and $N = 3, 5, 7$ in the polarized case in the Larin scheme calculated by the numerical 
codes. The comparison with analytic results using {\tt MATAD} 
\cite{Steinhauser:2000ry,Bierenbaum:2009mv} shows agreement by a relative error of better 
than 
$2 \cdot 10^{-11}$, with the exception of $(\Delta) 
A_{Qg}^{(3)}$, where the relative accuracy is better than $3 \cdot 10^{-8}$ in the unpolarized 
case  and $8 \cdot 10^{-9}$ in the polarized case. For $A_{gg}^{(3)}$ the moments are reproduced 
better than $3.6 \cdot 10^{-9}$.

\vspace*{4cm}
\begin{table}[H]\centering
\renewcommand*{\arraystretch}{1.4}
\begin{tabular}{|c|r|r|r|r|r|}
\hline
Parameter & $N_i$ & $a_i$ & $b_i$ & $c_i$ & $d_i$ \\
\hline
\multicolumn{1}{|c|}{} &  \multicolumn{5}{c|}{$Q^2 = 30~\GeV^2$} \\
\cline{2-6} 
$x\Delta^{\rm NS,+}$ & 0.10852 &  0.32896 & 3.0695 & 8.7302    & 0.21965 \\
$x\Sigma$      & 0.61061 & -0.30000 & 3.5000 & 5.0000    & 0.80000 \\
$xG$          & 1.17217 & -0.32230 & 6.0445 & 0.096178  & 0.00042125 \\
\hline
\multicolumn{1}{|c|}{} &  \multicolumn{5}{c|}{$Q^2 = 100~\GeV^2$} \\
\cline{2-6} 
$x\Delta^{\rm NS,+}$ & 0.09001 &  0.37333 & 3.2021 &  9.7870    & 0.12447 \\
$x\Sigma$        & 0.65237 & -0.32286 & 3.8537 &  5.1846    & 0.93115 \\
$xG$      & 2.14857 & -0.30370 & 2.8748 & -1.0000    & 0.23315 \\
\hline
\multicolumn{1}{|c|}{} &  \multicolumn{5}{c|}{$Q^2 = 10000~\GeV^2$} \\
\cline{2-6} 
$x\Delta^{\rm NS,+}$ & 0.063573 &  0.26936 & 3.7529 &  12.672    & 0.18461 \\
$x\Sigma$      & 0.726960 & -0.39337 & 6.2209 &  20.164    & 1.56240 \\
$xG$      & 11.86720 & -0.31917 & 3.5054 & -1.0000    & 0.03152 \\
\hline
\end{tabular}
\caption[]{\sf The effective parameterizations of the unpolarized parton distributions
at next-to-next-to-leading-order (NNLO).}
\label{TAB1}
\renewcommand*{\arraystretch}{1.0}
\end{table}

\newpage
\begin{table}[H]\centering
\renewcommand*{\arraystretch}{1.4}
{\footnotesize
\begin{tabular}{|r|r|r|}
\hline
OME                                           & unpolarized      & polarized     \\
\hline
$(\Delta){A}_{qq,Q,\delta}^{(0),\rm NS +}$    &  $1$             & $1$           
\\
$(\Delta){A}_{qq,Q,\delta}^{(0),\rm NS -}$    &  $1$             & $1$           
\\
$(\Delta){A}_{gg,Q,\delta}^{(0)}$             &  $1$             & $1$           
\\
\hline
$(\Delta){A}_{qq,Q,\delta}^{(0),\rm NS +}$    &  $1$             & $1$           
\\
$(\Delta){A}_{qq,Q,\delta}^{(0),\rm NS -}$    &  $1$             & $1$           
\\                                                                   
$(\Delta){A}_{Qg}^{(1)}$                      &  $-1.6400000000$ & $ 1.6000000000$ 
\\
$(\Delta){A}_{gg,Q,\delta}^{(1)}$             &  $ 1.6666666666$ & $ 1.6666666666$ 
\\
\hline
$(\Delta){A}_{qq,Q,\delta}^{(2),\rm NS +}$     &  $ 3.0170491475$ & $ 3.0170491475$ 
\\
$(\Delta){A}_{qq,Q,\delta}^{(2),\rm NS -}$     &  $ 3.0170491475$ & $ 3.0170491475$ 
\\
$(\Delta){A}_{qq,Q,+}^{(2),\rm NS +}$          &  $ 2.6951989026$ & $ 2.6951989026$ 
\\
$(\Delta){A}_{qq,Q,-}^{(2),\rm NS -}$          &  $ 2.6951989026$ & $ 2.6951989026$
\\
$(\Delta){A}_{qq,Q,\rm reg}^{(2),\rm NS +}$    &  $-1.4875157178$ & $-2.1071663281$
\\
$(\Delta){A}_{qq,Q,\rm reg}^{(2),\rm NS -}$    &  $-1.4875157178$ & $-2.1071663281$
\\
$(\Delta){A}_{Qq}^{(2),\rm PS}$                &  $-31.101625284$ & $ 3.3412446943$  
\\
$(\Delta){A}_{Qg}^{(2)}$                       &  $-65.439724185$ & $-1.4459037668$ 
\\
$(\Delta){A}_{gq,Q}^{(2)}$                     &  $ 43.210349430$ & $ 3.8739075190$ 
\\
$(\Delta){A}_{gg,Q,\delta}^{(2)}$              &  $ 3.0944444444$ & $ 3.0944444444$
\\
$(\Delta){A}_{gg,Q,+}^{(2)}$                   &  $ 6.0641975309$ & $ 6.0641975309$ 
\\
$(\Delta){A}_{gg,Q,\rm reg}^{(2)}$             &  $ 83.239813261$ & $ 2.4790593485$
\\                                                                   
\hline                                                            
$(\Delta){A}_{qq,Q,\delta}^{(3),\rm NS +}$     &  $-2.6422012017$ & $-2.6422012017$
\\
$(\Delta){A}_{qq,Q,\delta}^{(3),\rm NS -}$     &  $-2.6422012017$ & $-2.6422012017$ 
\\
$(\Delta){A}_{qq,Q,+}^{(3),\rm NS +}$          &  $-4.7428175642$ & $-4.7428175642$   
\\
$(\Delta){A}_{qq,Q,-}^{(3),\rm NS -}$          &  $-4.7428175642$ & $-4.7428175642$ 
\\
$(\Delta){A}_{qq,Q,\rm reg}^{(3),\rm NS +}$    &  $ 6.4686126472$ & $ 3.9315332719$ 
\\
$(\Delta){A}_{qq,Q,\rm reg}^{(3),\rm NS -}$    &  $ 6.6927772815$ & $ 4.0491109301$
\\
$(\Delta){A}_{Qq}^{(3),\rm PS}$                &  $ 91.950981088$ & $-39.972762346$
\\
$(\Delta){A}_{qq,Q}^{(3),\rm PS}$              &  $-38.624316410$ & $-1.7513673829$
\\
$(\Delta){A}_{Qg}^{(3)}$                       &  $ 310.17900321$ & $-99.957757769$
\\
$(\Delta){A}_{qg,Q}^{(3)}$                     &  $-73.710138327$ & $-6.7944302468$
\\
$(\Delta){A}_{gq,Q}^{(3)}$                     &  $-120.75198970$ & $-15.623565492$
\\
$(\Delta){A}_{gg,Q,\delta}^{(3)}$              &  $ 17.958634718$ & $ 17.958634718$   
\\
$(\Delta){A}_{gg,Q,+}^{(2)}$                   &  $-6.6285411655$ & $-6.6285411655$
\\
$(\Delta){A}_{gg,Q,\rm reg}^{(3)}$             &  $-262.65714922$ & $ 22.737595170$
\\
\hline
\end{tabular}}
\caption[]{\sf Cumulative test values for the single-mass OMEs $A_{ij}$, 
for $a_s = 1/10, \ln(m^2_Q/Q^2) = 10, x=1/10$ and $N_F = 3$ at zero, one, two and three 
loops in 
the unpolarized and polarized cases, supplemented by the weights of the 
distribution $\delta(1-x)$. 
In the polarized case we work in the Larin scheme. Vanishing terms are 
not listed.
}
\label{TAB2} 
\renewcommand*{\arraystretch}{1.0}
\end{table}
\begin{table}[H]\centering
\renewcommand*{\arraystretch}{1.4}
\begin{tabular}{|l|r|}
\hline
OME                                        & polarized    \\
\hline
$\Delta {A}_{qq,Q,NS,\delta}^{(0)}$        &  $1$    
\\
$\Delta {A}_{gg,Q,\delta}^{(0)}$           &  $1$      
\\
\hline
$\Delta {A}_{gg,Q,\delta}^{(1)}$           &  $-1.989703315309$              
\\
$\Delta {A}_{Qg}^{(1)}$                    &  $-1.989703315309$ 
\\
\hline                                          
$\Delta {A}_{qq,Q,\delta}^{(2),\rm NS}$    &  $ 13.25399267569$                 
\\
$ \Delta {A}_{qq,Q,+}^{(2),\rm NS}$        &  $ 5.520434456555$ 
\\
$\Delta {A}_{qq,Q,\rm reg}^{(2),\rm NS}$   &  $ 12.43689726767$ 
\\
$\Delta {A}_{Qq}^{(2),\rm PS}$             &  $-27.46655806489$ 
\\
$\Delta {A}_{Qg}^{(2)}$                    &  $-65.17404874351$ 
\\
$\Delta {A}_{gq,Q}^{(2)}$                  &  $ 25.07609255527$ 
\\
$\Delta {A}_{gg,Q,\delta}^{(2)}$           &  $-36.20966709533$ 
\\
$ \Delta {A}_{gg,Q,+}^{(2)}$               &  $ 12.42097752725$ 
\\
$ \Delta {A}_{gg,Q,reg}^{(2)}$             &  $ 7.833545304545$ 
\\
\hline
${A}_{qq,Q,\delta, \rm trans}^{(2)}$       &  $ 13.25399267569$ 
\\
${A}_{qq,Q,+,\rm trans}^{(2)}$             &  $ 5.520434456555$ 
\\
${A}_{qq,Q,reg,\rm trans}^{(2)}$           &  $ 1.650250988934$ 
\\
\hline
\end{tabular}
\caption[]{\sf Individual test values for the different contributions in $a_s$ to the 
polarized single-mass OMEs $\Delta A_{ij}$ and the ones for transversity in the 
$\overline{\rm MS}$ scheme up to
two-loop order for $x = 1/3, Q^2 = 50~\GeV^2$ and $m_c = 1.59~\GeV$, supplemented by the 
weights of the distribution $\delta(1-x)$.
}
\label{TAB3} 
\renewcommand*{\arraystretch}{1.0}
\end{table}

\begin{table}[H]\centering
\renewcommand*{\arraystretch}{1.4}
\begin{tabular}{|r|r|r|r|}
\hline
\multicolumn{1}{|c|}{OME} & \multicolumn{3}{c|}{moments}\\
\cline{2-4}
\multicolumn{1}{|c|}{   } & \multicolumn{1}{c|}{$N=2$} & \multicolumn{1}{c|}{$N=4$}
& \multicolumn{1}{c|}{$N=6$} \\
\hline
$A_{qq,Q}^{\rm NS, even}$ &   2.552050805 &  5.205091550 &  6.935755383 \\ 
$A_{Qq}^{\rm PS}$         &   5.787728157 &  1.517439651 &  0.726319249 \\
$A_{qq,Q}^{\rm PS}$       & --1.927340435 &--0.287734919 &--0.111383030 \\
$A_{Qg}$                  & --4.267201565 &--19.75022897 &--21.85267609 \\
$A_{qg,Q}$                &   2.466090032 &  3.364276844 &  2.718842618 \\       
$A_{gq,Q}$                & --5.412438528 &--0.735206982 &--0.219958419 \\ 
$A_{gg,Q}$                &   2.801111705 &  20.85194977 &  26.04572674 \\       
\hline
\multicolumn{1}{|c|}{   } & \multicolumn{1}{c|}{$N=3$} & \multicolumn{1}{c|}{$N=5$}
& \multicolumn{1}{c|}{$N=7$} \\
\hline
$A_{qq,Q}^{\rm NS, odd}$  &   4.041765636 &  6.148438117 &  7.610661858 \\
\hline
\end{tabular}
\caption[]{\sf Test values for the moments of the unpolarized OMEs $A_{ij}$ 
for $a_s = 1/10, \ln(m^2_Q/Q^2) = 10, N_F=3$.}
\label{TAB4}
\renewcommand*{\arraystretch}{1.0}
\end{table}

\begin{table}[H]\centering
\renewcommand*{\arraystretch}{1.4}
\begin{tabular}{|r|r|r|r|}
\hline
\multicolumn{1}{|c|}{OME} & \multicolumn{3}{c|}{moments}\\
\cline{2-4}
\multicolumn{1}{|c|}{   } & \multicolumn{1}{c|}{$N=3$} & \multicolumn{1}{c|}{$N=5$}
& \multicolumn{1}{c|}{$N=7$} \\
\hline
$\Delta A_{qq,Q}^{\rm NS, odd}$  &   3.768279839 &  6.034765595 &  7.548388996 \\ 
$\Delta A_{Qq}^{\rm PS}$         &   0.925763149 &  0.605732786 &  0.388711691 \\
$\Delta A_{qq,Q}^{\rm PS}$       & --0.352600132 &--0.135250183 &--0.068551252 \\
$\Delta A_{Qg}$                  & --12.00694753 &--18.80322492 &--20.64374519 \\
$\Delta A_{qg,Q}$                &   2.844204630 &  2.667371162 &  2.296278935 \\       
$\Delta A_{gq,Q}$                & --0.720767851 &--0.130314549 &  0.022959543 \\ 
$\Delta A_{gg,Q}$                &   16.74536513 &  23.16909172 &  26.75182876 \\       
\hline
\multicolumn{1}{|c|}{   } & \multicolumn{1}{c|}{$N=2$} & \multicolumn{1}{c|}{$N=4$}
& \multicolumn{1}{c|}{$N=6$} \\
\hline
$\Delta A_{qq,Q}^{\rm NS, even}$ &   2.030572581 &  5.039758718 &  6.854626520 \\
\hline
\end{tabular}
\caption[]{\sf Test values for the moments of the polarized OMEs $\Delta 
A_{ij}$ for $a_s = 1/10, \ln(m^2_Q/Q^2) = 10, N_F=3$ in the Larin scheme.}
\label{TAB5}
\renewcommand*{\arraystretch}{1.0}
\end{table}
\section{Numerical Codes}
\label{APP_B}

\vspace*{1mm}
\noindent
In the following we describe the numerical implementations of the massive OMEs 
in $x$-space contained in ancillary files to this paper.

We provide a {\tt C++} code {\tt libome} for the numerical representation of the OMEs 
in terms of precise local overlapping series expansions for the unpolarized and 
polarized massive OMEs to three-loop order. 
In addition, the code can be accessed at {\tt https://gitlab.com/libome/libome}.
For even faster implementations, the {\tt 
Fortran} codes {\tt OMEUNP3} for the unpolarized case and {\tt OMEPOL3}
for the polarized case are provided. These codes shall be stored in 
different directories and can be compiled by using {\tt gfortran}.
The codes cover the OMEs in the region
$x \in [10^{-6},1]$. They are based on small and large-$x$ expansions
and precise grids describing the remainder parts, with validity for $x \in 
[10^{-6},1]$. For the remainder part we use 
cubic spline interpolation \cite{SPLINE}.\footnote{We thank S. Kumano and M. Miyama of 
the AAC-collaboration for allowing us to use their interpolation
routines.} 500 points are 
logarithmically distributed in the low-$x$ region and linearly in the region of larger 
values of $x$.
The spline tables need to be calculated only once. This is done by setting {\tt IGRID = 
0} through which the spline parameter are written to disk. 
Subsequently, by setting {\tt IGRID = 1}, the tables can be 
read in from disk to allow for fastest possible performance of the code.
Before calculating the spline parameters newly, the file {\tt UOUT.dat} has to be 
removed.  
The OMEs are presented as 
polynomials in $a_s, N_F$ and $\ln(m^2_Q/Q^2)$, which is necessary for the fits.
Furthermore, the codes can be run at tree-level $(N=0)$, taking all terms up to one loop 
$(N=1)$, up to two loops $(N=2)$, and up to three loops $(N=3)$.
These {\tt Fortran} codes are particularly suited for fitting codes.

In earlier versions, we have used harmonic polylogarithm 
\cite{Gehrmann:2001pz,Ablinger:2018sat} and Nielsen integral representations
\cite{NIELSEN1,Kolbig:1983qt,Devoto:1983tc,LEWIN1,LEWIN2,Blumlein:2000hw}.
These are, however, slower than the present codes. Up to two-loop order we provide 
the code {\tt OMEPOL2MS} based on this in the polarized case in the 
$\overline{\rm MS}$ scheme.
 
\section{\boldmath The constant part $\Delta a_{ij}$ of some unrenormalized OMEs}
\label{APP_C}

\vspace*{5mm} 
\noindent
In past publications we have illustrated all single mass OMEs $(\Delta) A_{ij}^{(3)}$
or the constant parts of the respective unrenormalized QMEs, $(\Delta) a_{ij}^{(3)}$, 
in $x$-space in Refs.~\cite{Ablinger:2014vwa,
Ablinger:2014nga,
Ablinger:2022wbb,
Ablinger:2023ahe,
Ablinger:2024xtt} and \cite{Ablinger:2014nga,Ablinger:2019etw,Ablinger:2022wbb,
Ablinger:2023ahe,
Ablinger:2024xtt}, 
but not yet for $(\Delta) a_{qg,Q}^{(3)},  
(\Delta) a_{qq,Q}^{(3),\rm PS}$ \cite{Behring:2014eya,Blumlein:2021xlc} and
$(\Delta) a_{gq,Q}^{(3)}$ \cite{Ablinger:2014lka,Behring:2021asx}. We show these for completeness
in the following. The terms $(\Delta) a_{ij}(x)$ are the most involved parts of the respective 
massive OMEs to three-loop order.
In the illustrations we set $N_F=3$. In the small-$x$ region we always expand to the constant term,
as well as in the large-$x$ 
limit. 

\begin{figure}[H]
\centering
\includegraphics[width=0.49\textwidth]{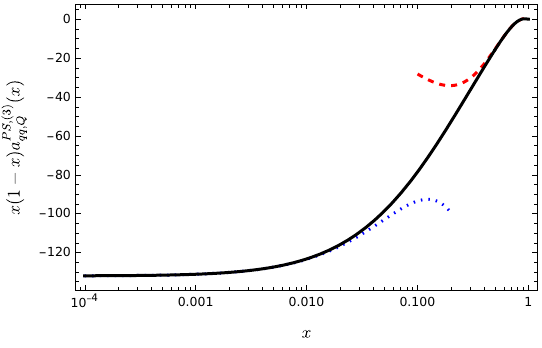}
\includegraphics[width=0.49\textwidth]{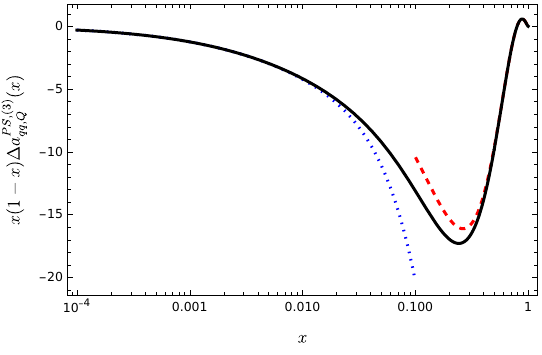}
\caption{\sf 
Full lines: the distributions $(\Delta) a_{qq,Q}^{\rm PS, (3)}(x)$. Dotted lines: small $x$ expansion 
up to the constant term. Dashed lines:  large $x$ expansion up to terms of $O((1-x)^5)$. 
\label{fig:11}}
\end{figure}

\begin{figure}[H]
\centering
\includegraphics[width=0.49\textwidth]{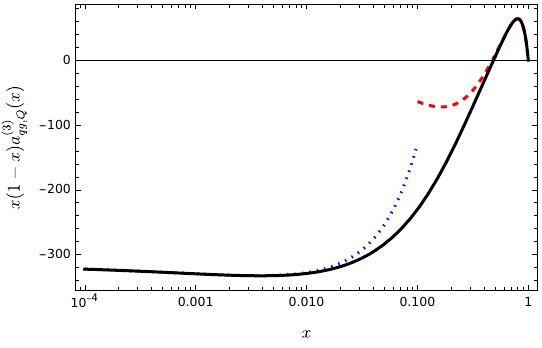}
\includegraphics[width=0.49\textwidth]{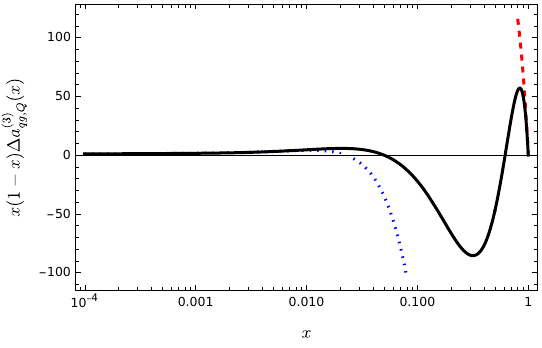}
\caption{\sf  
Full lines: the distributions $(\Delta) a_{qg,Q}^{(3)}(x)$. Dotted lines: small $x$ expansion 
up to the constant term. Dashed lines:  large $x$ expansion in the unpolarized case: constant term;
in the polarized case to $(1-x)^5$.
\label{fig:12}}
\end{figure}

\begin{figure}[H]
\centering
\includegraphics[width=0.49\textwidth]{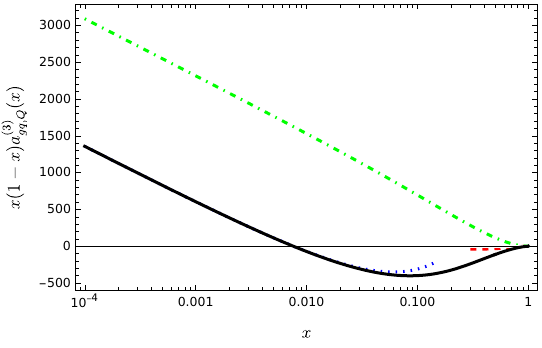}
\includegraphics[width=0.49\textwidth]{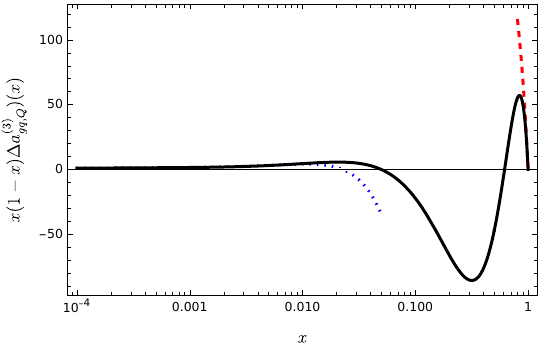}
\caption{\sf 
Full lines: the distributions $(\Delta) a_{gq,Q}^{(3)}(x)$. Dotted lines: small $x$ expansion 
up to the constant term. Dashed lines:  large $x$ expansion to the constant term. Dash-dotted line for 
$a_{gq,Q}^{(3)}(x)$: leading small-$x$ term of $O(\ln(x)/x)$.
\label{fig:13}}
\end{figure}

Figure~\ref{fig:11} depicts $a_{qq,Q}^{(3),\rm PS}$ and $(\Delta) a_{qq,Q}^{(3),\rm PS}$.
We show also the small-$x$ and large-$x$ expansion terms. 
Here it has been necessary to expand to 
terms 
of $O((1-x)^5)$ in some cases to show the agreement in a wider range in the large 
$x$-region.
In Figure~\ref{fig:12} $a_{qg,Q}^{(3)}$ and $(\Delta) a_{qg,Q}^{(3)}$ are shown.
Also here the large-$x$ expansion in the polarized case had to be extended to terms of 
$O((1-x)^5)$.
Finally, in Figure~\ref{fig:13}, $a_{gq,Q}^{(3)}(x)$ and $\Delta a_{gq,Q}^{(3)}(x)$
are shown. As an example, we depict here also the leading small-$x$ term for $a_{gq,Q}^{(3)}(x)$.
As has been demonstrated in many other cases before, see e.g. Refs.~\cite{Blumlein:1997em,
Blumlein:1998mg,Blumlein:1999ev,Ablinger:2014nga,Ablinger:2022wbb,Ablinger:2024xtt},
this term is not describing $a_{gq,Q}^{(3)}(x)$ anywhere due to compensating effects of
subleading contributions, which are large.

{\bf Acknowledgment.} We thank S.~Klein and A.~Goedicke for the calculation of 
a number of polarized Mellin moments in earlier collaboration, cf.~Ref.~\cite{SKLEIN}. 
We would like to thank H.~B\"ottcher, P.~Marquard and M.~Saragnese for discussions. 
This work was supported by the European Research Council (ERC)
under the European Union's Horizon 2020 research and innovation programme
grant agreement 101019620 (ERC Advanced Grant TOPUP) and the UZH Postdoc Grant,
grant no.~[FK-24-115] and by the Austrian Science Fund (FWF) Grant-DOI 10.55776/P20347. 
We thank the Inspire team for their enormous work \cite{Crepe-Renaudin:2025rck} without 
of which a proper literature search and document retrieval for daily scientific work 
would be completely impossible.

{\small

}
\end{document}